\documentclass[aps,prl,twocolumn,showpacs,superscriptaddress]{revtex4-1}
\usepackage{amsmath,amssymb,graphicx,color}
\usepackage{times}
\begin{document}

\newcommand\ket[1]{\left|\textstyle{#1}\right\rangle}
\newcommand\bra[1]{\left\langle\textstyle{#1}\right|}
\newcommand\braket[1]{\left\langle\textstyle{#1}\right\rangle}
\newcommand\half{\frac{1}{2}}
\newcommand\red{\color{red}}

\title{Large-Scale Schr\"odinger-Cat States and Majorana Bound States in Coupled Circuit-QED Systems}

\author{Myung-Joong Hwang}
\affiliation{Department of Physics, Pohang University of Science and
  Technology, Pohang 790-784, Korea}

\author{Mahn-Soo Choi}
\affiliation{Department of Physics, Korea University, Seoul 136-713,
  Korea}

\begin{abstract}
We have studied the low-lying excitations of a chain of coupled circuit-QED
systems in the ultrastrong coupling regime, and report several intriguing
properties of its two nearly degenerate ground states. The ground states are
Schr\"odinger cat states at a truly large scale, involving maximal
entanglement between the resonators and the qubits, and are mathematically
equivalent to Majorana bound states. With a suitable design of physical
qubits, they are protected against local fluctuations and constitute a
non-local qubit. Further, they can be probed and manipulated coherently by
attaching an empty resonator to one end of the circuit-QED chain.
\end{abstract}
\maketitle

Confronted with formidable difficulties in solving strongly interacting
many-body systems, it has been desired to find good quantum simulators.  It
may seem natural to simulate a many-body system with another tunable system of
massive particles such as ultracold atomic gases~\cite{Bloch:2012jy}.  In
fact, any controllable quantum system, notably quantum computer if ever
practical, can simulate efficiently many-body systems~\cite{Feynman:1982gn}.
Indeed it has been recognized that photons confined in
coupled-cavities simulate closely the quantum behaviors of strongly-correlated
many-body
systems~\cite{Hartmann:2006kv,*Greentree:2006jg,Angelakis:2007ho}. Subsequent
studies have revealed that Bose-Hubbard
model~\cite{Rossini:2007fx,*Irish:2008ci,*Makin:2008hr,*Koch:2009hh,*Schmidt:2009cs,*Schmidt:2010kl},
interacting spin
models~\cite{Angelakis:2007ho,Hartmann:2007db,*Kay:2008ip,*Cho:2008et}, and
other exotic quantum phases~\cite{Cho:2008ct,*Carusotto:2009dr,*Koch:2010eu}
can be simulated efficiently using the coupled-cavities.
Further, recent advances in solid-state devices such as circuit-QED
systems~\cite{Wallraff:2004dy,Schoelkopf:2008cs} and
micro-cavities~\cite{Aoki:2006gq,Hennessy:2007hi} and ongoing efforts to
fabricate large-scale cavity arrays~\cite{Houck:2012iq,*Underwood:2012vq} make
the array of coupled cavities a promising candidate for an efficient quantum
simulator.

Meanwhile, the ultrastrong coupling regime of the cavity-QED system, where
the light-matter coupling energy is comparable to or even higher than the
energy of the cavity field, has been
envisioned~\cite{Devoret:2007gi,*Bourassa:2009gy,*Bourassa:2012tv} and
experimentally
demonstrated~\cite{Niemczyk:2010gv,*FornDiaz:2010by,*Gunter:2009gc}.  The
ultrastrong coupling brings about fundamentally different physics deeply
connected to the high degree of entanglement between the ``matter'' and the
photon~\cite{Ashhab:2010eh,Hwang:2010jn,Nataf:2010dy,Nataf:2011ff,Braak:2011hc,*Irish:2007bo,*Hausinger:2010eb,*Hausinger:2011bc,*Casanova:2010kd,*Zueco:2009it}.
However, the effect of ultrastrong coupling on the low-energy excitations of
an array of coupled cavity-QED systems remains unclear, and is our main
concern in this work.

\begin{figure}[b]
\centering
\includegraphics[width=8cm]{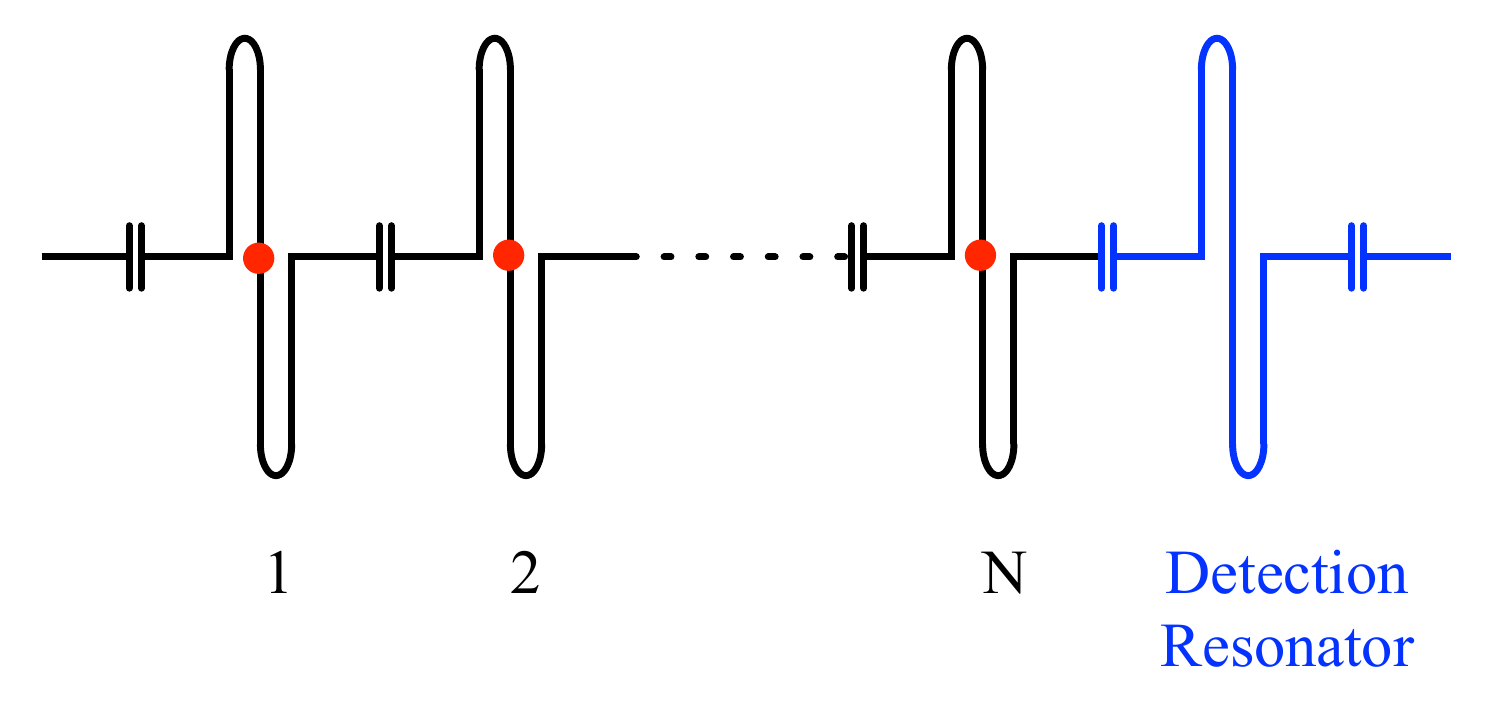}
\caption{Schematic of 1D circuit-QED arrays. The red dots indicate
  qubits placed inside of superconducting resonator. The $N$th
  resonator is coupled to the detection resonator. The circuit-QED
  array realizes the transverse field Ising model (TFIM), and the detection
  resonator can measure and control the degenerate ground state of the
  TFIM.}
\label{fig:1}
\end{figure}

In this paper, we investigate the low-lying excitations of a one-dimensional
(1D) array of circuit-QED systems (cQEDs), with each cQED being in the
ultrastrong coupling regime; see Fig.~\ref{fig:1}.
It turns out that the array permits two nearly degenerate ground states
separated by a finite energy gap from the continuum of higher-energy states.
We find several intriguing properties of the two ground states: (i) They are
Schr\"odinger cat states at a truly large scale, and involve maximal
entanglement between the resonators and the qubits. (ii) With a suitable
design of physical qubits, the two ground states are protected against local
fluctuations and constitute a non-local qubit~\cite{Tserkovnyak:2011dl}. (iii)
They are mathematically equivalent to the long-searched Majorana bound
states~\cite{[To be compared with~]Bardyn:2012td}.  (iv) They can be probed
and manipulated coherently by attaching an empty resonator to one end of the
circuit-QED chain. Such configuration turns the total system (the circuit-QED
chain plus the empty resonator) into another effective circuit-QED system.
There are many promising types of superconducting qubits, among which we focus
on Fluxonium~\cite{Manucharyan:2009fo,*Koch:2009ec}. As we illustrate below,
its strong inductive coupling with the superconducting
resonator~\cite{Nataf:2010dy} and its anisotropic noise
characteristics~\cite{Nataf:2011ff} are well suited for our purpose.

\emph{System: a circuit-QED chain} --- We consider a 1D array of cQEDs; see
Fig.~\ref{fig:1}. Each cQED consists of the ``resonator'', a superconducting
microwave transmission line, and the ``qubit'', a superconducting quantum bit
(two-level system)~\cite{Blais:2004kn,Wallraff:2004dy}, and is theoretically
described by the Rabi Hamiltonian
\begin{equation}
\label{paper::eq:2}
H^\mathrm{cQED}_i = \omega_0 a_i^\dagger a_i
-\lambda(a_i+a_i^\dagger)\sigma_i^x+\frac{\Omega}{2}\sigma_i^z
\end{equation}
where $a_i$ and $a_i^\dag$ are the field operators of the resonator with
frequency $\omega_0$, $\sigma_i^x$ and $\sigma_i^z$ Pauli operators of the
qubit with energy splitting $\Omega$, and $\lambda$ the resonator-qubit
coupling energy in the $i$th cQED.
The resonators of neighboring cQEDs are coupled capacitively to each other,
and photons hop from one resonator to nearby ones.  The
Hamiltonian of the whole chain is thus given by
\begin{equation}
\label{paper::eq:1}
H=\sum_i^N H^\mathrm{cQED}_i -J\sum_{i}^{N-1}
(a^\dagger_i a_{i+1}+ a_{i}a^\dagger_{i+1})
\end{equation}
where $J$ is the photon hopping amplitude and $N$ is the number of cQEDs
in the chain.

Before discussing the energy levels and associated wavefunctions of the whole
chain, we first briefly review the properties of the low-lying states of a
single cQED in the ultrastrong coupling regime ($\lambda\gtrsim\omega_0$). The
strong coupling disables the standard rotating wave approximation, which
reduces Eq.~(\ref{paper::eq:2}) to the Jaynes-Cummings Hamiltonian.
As a consequence, the ground state of Eq.~(\ref{paper::eq:2}) is not a simple
vacuum anymore as in the Jaynes-Cumming model. Instead, it contains finite
average photon numbers, and shows non-classical properties such as squeezing
and entanglement~\cite{Ashhab:2010eh,Hwang:2010jn}.
To see this, let us examine the ground-state wavefunction more closely:
Approximate expressions for the nearly-degenerate ground states have been
derived in Ref.~\onlinecite{Nataf:2010dy} (see also
Ref.~\onlinecite{Hwang:2010jn}).  Here we take a different approach and
explore the parity symmetry in the Rabi Hamiltonian, which is important to
understand the effect of photon hopping. The Hamiltonian (\ref{paper::eq:2})
commutes with the ``parity'' operator $\Pi_i=\exp(-i\pi a_i^\dagger
a_i)\sigma_i^z$, and thus the Hilbert space is classified into subspaces
$\mathcal{E}_i^\pm$ of $\pm$ parity. Within each subspace $\mathcal{E}_i^\pm$,
the Hamiltonian can be described in effect by a single bosonic operator,
$b_i=a_i\sigma^x_i$:
\begin{math}
H_i^\mathrm{cQED}\to H^\pm_i=H_i^0\pm H_i^1
\end{math}
with
\begin{math}
H^0_i =\omega_0(b_i^\dagger-\lambda/\omega_0)
(b_i-\lambda/\omega_0)-\lambda^2/\omega_0
\end{math}
and $H^1_i=\frac{\Omega}{2}\cos(\pi b_i^\dagger
b_i)$~\cite{Hwang:2010jn}. $H^0_i$ is simply a displaced harmonic oscillator
and the ground state is a coherent state $\ket{\lambda/\omega_0}_{b_i}^\pm$.
For $\lambda/\omega_0\gg1$ (regardless of $\Omega$), $H^1_i$ can be treated
perturbatively and shifts the energies of $\ket{\lambda/\omega_0}_{b_i}^\pm$
relatively by an exponentially small amount $\Delta=\frac{\Omega}{2}
e^{-2(\lambda/\omega_0)^2}$.
Now, back in the $\{a_i,\sigma_i^z\}$-basis, the nearly degenerate
ground states $\ket{\lambda/\omega_0}_{b_i}^\pm$ are expressed as
\begin{subequations}
\label{paper::eq:3}
\begin{align}
\label{paper::eq:3a}
\ket{0}_i&\equiv \frac{1}{\sqrt{2}}\left(\ket{\lambda/\omega_0}_i\ket{+}_i-\ket{-\lambda/\omega_0}_i\ket{-}_i\right), \\
\label{paper::eq:3b}
\ket{1}_i&\equiv \frac{1}{\sqrt{2}}\left(\ket{\lambda/\omega_0}_i\ket{+}_i+\ket{-\lambda/\omega_0}_i\ket{-}_i\right),
\end{align}
\end{subequations}
where $\ket{\alpha}_i$ ($\alpha\in\mathbb{C}$) is the eigenstate (coherent
state) of $a_i$ and $\ket{\pm}_i$ are the eigenstates of $\sigma_i^x$.
In short, these two ground states, $\ket{0}_i$ and $\ket{1}_i$, residing in
distinct parity subspaces are nearly degenerate with an energy splitting of
$2\Delta$, separated far from higher-energy states by an energy gap
$\omega_0$.

\emph{Effective model: a transverse-field Ising chain} --- Let us now
investigate the whole chain described by the Hamiltonian
(\ref{paper::eq:1}). Typically $J\ll\omega_0$, and we are mainly interested in
the low-lying excitations, well below $\omega_0$. In this limit, each cQED
remains within the subspace spanned by the states $\ket{0}_i$ and $\ket{1}_i$
in Eq.~(\ref{paper::eq:3}) and can be regarded as a \textit{pseudo-spin}:
\begin{equation}
\label{paper::eq:4}
\sum_i^N H^\mathrm{cQED}_i= -\Delta\sum_i^N \tau^z_i
\end{equation}
where $\tau^z_i=\ket{0}_i\bra{0}-\ket{1}_i\bra{1}$ and the energy splitting
$\Delta$ plays the role of Zeeman field. Hopping of a photon into or out of a
cavity changes the parity of its state, or more explicitly
$a_i\ket{0}_i=\lambda/\omega_0\ket{1}_i$ and
$a_i\ket{1}_i=\lambda/\omega_0\ket{0}_i$. Based on these observation, we can
identify $a_i$ and $a^\dagger_i$ as a pseudo-spin-flip operator $\tau^x_i$ and
$a_ia^\dagger_{i+1}$ as Ising interaction $\tau^x_i\tau^x_{i+1}$. That is, the
photon-hopping part of the Hamiltonian becomes
\begin{equation}
\label{paper::eq:6}
J\sum_{i}^{N-1} (a^\dagger_i a_{i+1}+ a_{i}a^\dagger_{i+1})= J_\mathrm{eff}\sum_i^{N-1} \tau^x_i\tau^x_{i+1}
\end{equation}
with $J_\mathrm{eff}=2J(\lambda/\omega_0)^2$.
The effective Ising interaction strength, $J_\mathrm{eff}$, is renormalized
with respect to $J$ by the factor $(\lambda/\omega_0)^2$ because the field
part of the pseudo-spin states in Eq.~(\ref{paper::eq:3}) is a coherent
state with amplitudes $\lambda/\omega_0$ and the field-field interaction
between resonators is proportional to the amplitudes of the resonator fields.

Putting both terms in Eqs.~(\ref{paper::eq:4}) and (\ref{paper::eq:6})
together, the low-energy effective Hamiltonian for the cQED chain becomes the
so-called transverse-field Ising model (TFIM),
\begin{equation}
\label{paper::eq:7}
H_\mathrm{Ising}
= -\Delta\sum_i^N \tau^z_i-J_\mathrm{eff}\sum_i ^{N-1}\tau^x_i\tau^x_{i+1}.
\end{equation}
The TFIM exhibits a quantum phase transition between the \emph{magnetically
  ordered} phase for $\Delta<J_\mathrm{eff}$ and the \emph{quantum paramagnet}
phase for $\Delta>J_\mathrm{eff}$~\cite{Sachdev:2011uj}.  The former is
particularly interesting for our purposes. For $\Delta=0$, $H_\mathrm{Ising}$
has two degenerate ground states,
$\ket{\Rightarrow}\equiv\prod_i\ket{\rightarrow}_i$ and
$\ket{\Leftarrow}\equiv\prod_i\ket{\leftarrow}_i$, where $\ket{\rightarrow}_i$
and $\ket{\leftarrow}_i$ are eigenstates of $\tau_i^x$. For $\Delta>0$ (yet
$\Delta<J_\mathrm{eff}$), $\tau_i^z$ tends to flip the pseudo-spins,
$\ket{\rightarrow}_i\leftrightarrow\ket{\leftarrow}_i$. It causes tunneling
between $\ket{\Rightarrow}$ and $\ket{\Leftarrow}$ via soliton propagation,
and hence the true eigenstates become
\begin{equation}
\label{paper::eq:10}
\ket{\Psi_0}
= \frac{1}{\sqrt{2}}\left(\ket{\Rightarrow}+\ket{\Leftarrow}\right)
\,,\quad
\ket{\Psi_1}
= \frac{1}{\sqrt{2}}\left(\ket{\Rightarrow}-\ket{\Leftarrow}\right)
\end{equation}
However, as the tunneling involves $N$ spins, the tunneling amplitude is
exponentially suppressed with the system size $N$. In other words,
$\ket{\Psi_0}$ and $\ket{\Psi_1}$ are nearly degenerate with energy splitting,
$\delta\sim\exp(-N/\xi)$ with $\xi$ being the correlation length of the Ising
chain, exponentially small in system size $N$. Both are separated from the
continuum of excitations by the energy gap $J_\mathrm{eff}$.

The two states $\ket{\Psi_0}$ and $\ket{\Psi_1}$ in Eq.~(\ref{paper::eq:10})
have non-local combinations of many pseudo-spins and are widely known as
Greenberger-Horne-Zeilinger (GHZ) states \cite{Greenberger89a}.  Moreover, by
expressing them in the original $\{a_i,\sigma_i^x\}$-basis
\begin{equation}
\label{paper::eq:12}
\ket{\Psi_s} =
\frac{1}{\sqrt{2}}
\left[\prod_i^N\ket{\lambda/\omega_0}_i\ket{+}_i
  +(-1)^s\prod_i^N\ket{-\lambda/\omega_0}_i\ket{-}_i\right]
\end{equation}
with $s=0$ or $1$, one can see that they involve high degree of non-local
entanglement between cavity fields and qubits. They are thus
\emph{Schr\"odinger cat states} at a truly large scale while many
theoretically proposed or experimentally demonstrated Schr\"odinger cat
states~\cite{Yurke:1986cm,Ourjoumtsev:2007cw} contain merely a single
radiation field. Below we illustrate that the two states in
(\ref{paper::eq:12}) are protected against local fluctuations and constitute a
non-local qubit~\cite{Tserkovnyak:2011dl}.

\emph{Effective model: a Majorana chain} --- 1D TFIM discussed above is
equivalent to a chain of Majorana fermions
\cite{Kitaev:2001up,Kitaev:2009ut}. The latter has attracted great interest
because it permits localized Majorana modes that can be used for topologically
protected quantum computation
\cite{Kitaev:2001up,Kitaev:2009ut,Kitaev:2006ik}. A very recent
experiment~\cite{Mourik:2012je} suggests that the Majorana chain can be
realized in a solid-state system, and intensive efforts are made in this
direction~\cite{[{For a recent review, see }]Alicea:2012wg}.

Here we re-express the two nearly degenerate states in
Eq.~(\ref{paper::eq:10}) or (\ref{paper::eq:12}) in terms of localized
Majorana fermions, and later discuss an experimentally feasible way of probing
such Majorana fermions.
The equivalence between the TFIM and the Majorana chain can be seen through a
Jordan-Wigner transformation~\cite{Lieb:1961dn}:
$c^\dagger_i=\tau_i^+\prod^{i-1}_{j=1}(-\tau^z_j)$ with
$\tau^+_i=\half(\tau^x_i+i \tau^y_i)$.  The operators $c_i$ and $c_i^\dag$
describe Dirac fermions and satisfy $\{c_i,c^\dagger_j\}=\delta_{ij}$ and
$\{c_i,c_j\}=0$. The Dirac fermion operators are further represented with
self-conjugate Majorana operators, $\gamma_{2i-1}=c^\dagger_i+c_i$ and
$\gamma_{2i}=i(c^\dagger_i-c_i)$. The TFIM (\ref{paper::eq:7}) is then reduced
to
\begin{equation}
\label{paper::eq:8}
H_\mathrm{Majorana} = \frac{i}{2}\left[
  \Delta\sum_{i=1}^N\gamma_{2i-1}\gamma_{2i}+
  J_\mathrm{eff}\sum_i^{N-1}\gamma_{2i}\gamma_{2i+1}\right]
\end{equation}
At $\Delta=0$, the Majoranas at the two ends, $\gamma_1$ and $\gamma_{2N}$, in
the chain does not appear in the Hamiltonian, which implies the existence of two
degenerate ground states. These are nothing but $\ket{\Rightarrow}$ and
$\ket{\Leftarrow}$ in Eq.~(\ref{paper::eq:10}).
For finite $\Delta$, the two states $\ket{\Rightarrow}$ and $\ket{\Leftarrow}$
are mixed linearly into $\ket{\Psi_0}$ and $\ket{\Psi_1}$ in
Eq.~(\ref{paper::eq:10}) due to the tunneling between two Majorana modes
$\gamma_1$ and $\gamma_{2N}$, and the degeneracy is lifted. Since the
tunneling is through the whole chain, the energy splitting $\delta$ is
exponentially small (as long as $\Delta<J_\mathrm{eff}$).
One can check that $(\gamma_{1}+i \gamma_{2N})\ket{\Psi_0} = 0$ and
$(\gamma_{1}+i \gamma_{2N})\ket{\Psi_1} = 2\ket{\Psi_0}$, which means that
$\ket{\Psi_1}$ has one more fermion than $\ket{\Psi_0}$ or equivalently that
$\ket{\Psi_0}$ and $\ket{\Psi_1}$ have different fermion parities.

Here we emphasize that the two Majoranas localized at the ends of the Majorana
chain are actually non-local in the physical chain, i.e., the cQED chain or the
Ising chain~\cite{Bardyn:2012td}: The Majorana operators is represented in
terms of $\tau_j^x$ and $\tau_j^z$ as
\begin{equation}
\label{paper::eq:9}
\gamma_{1}=\tau_1^x,\quad
\gamma_{2N}=i\tau^x_N \prod_{j=1}^{N}(-\tau^z_j),
\end{equation}
and $\gamma_{2N}$ involves the \emph{string} operator
$\prod_{j=1}^{N}(-\tau^z_j)$.  This implies that the two nearly-degenerate
ground states $\ket{\Psi_0}$ and $\ket{\Psi_1}$ are not protected
topologically against local noise even though mathematically they correspond
to two distinct Majorana modes.  It is in stark contrast to the case where the
two Majorana modes at the ends of a $p$-wave superconducting wire are
topologically protected. However, we will see below that the two
states $\ket{\Psi_0}$ and $\ket{\Psi_1}$ are vulnerable only to a certain type
of local noise and there exist realistic systems with such type of local noise
significantly suppressed.

\emph{Noise} --- It is evident from the expression in Eq.~(\ref{paper::eq:12})
that the non-local spin qubits are prone to the local noise in $\sigma_i^x$ of
the physical qubits and the one in $a_i+a_i^\dagger$ of the resonators.  The
states are intrinsically robust against the $\sigma_i^y$ and $\sigma_i^z$ noise
since ${}_i{\bra{\Psi_1}}\sigma_i^{y,z}\ket{\Psi_0}_i\sim
e^{-(\lambda/\omega_0)^2}$, which is reminiscent of the Franck-Condon effect.
The $a_i+a_i^\dag$ noise affects only the resonator at the end of the chain
(which is usually connected external microwave environment for measurement),
and can be easily avoided by replacing it by a high-Q resonator.
The problem with $\sigma_i^x$ noise can be circumvented, for example, by using
Fluxonium for qubits. Fluxonium is known to have anisotropic noise
characteristics with $\sigma_i^{y,z}$ being the dominant noises and the
$\sigma_i^x$ noise ignorable~\cite{Nataf:2011ff}.

What about the effect of inhomogeneity in system parameters?  Above we have
assumed $\omega_0$, $\Omega$, $\lambda$ and $J$ of each circuit-QED to be
homogeneous. Deviations in $\omega_0$, $\Omega$, $\lambda$ lead to
fluctuations in $\Delta$. The inter-cavity coupling strength, $J$, can also be
varied from cavity to cavity, which leads to inhomogeneous TFIM,
\begin{equation}
-\sum_i\Delta_i\tau_i^z -\sum_i J_i^\mathrm{eff}\tau_i^x\tau_{i+1}^x.
\end{equation}
This Hamiltonian still conserves the parity symmetry,
$\mathcal{P}=\prod^N_{i=1}\tau_z^i$ which is respected by the degenerate
ground states. Therefore, the ground states will be robust to small
fluctuations in $\Delta_i$ and $J_i$.
We thus conclude that the nearly degenerate ground states $\ket{\Psi_0}$ and
$\ket{\Psi_1}$ can be kept well protected by a careful design of the physical
qubits in the system.

\begin{figure}
\centering
\includegraphics[width=8cm]{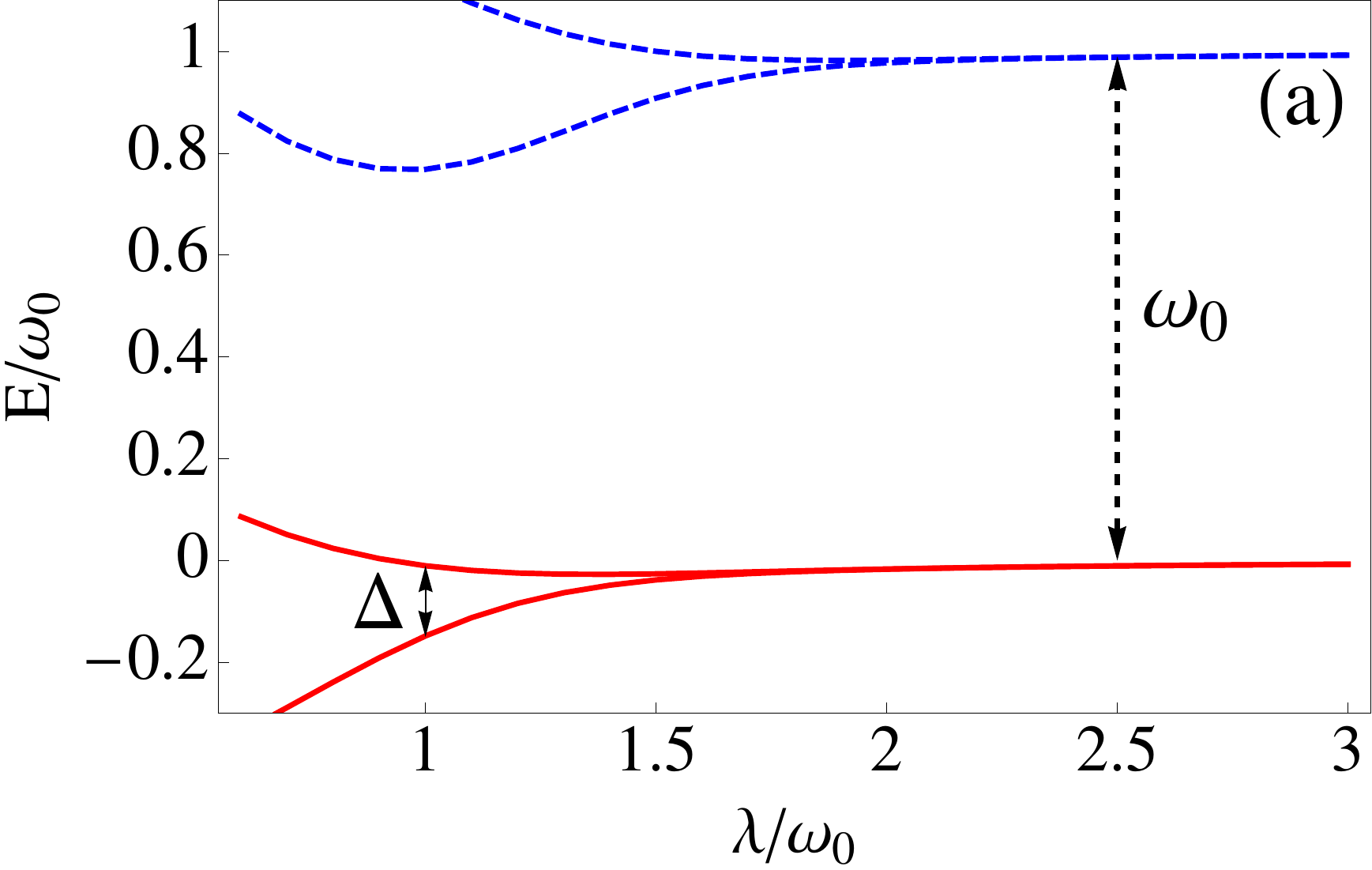}
\includegraphics[width=8cm]{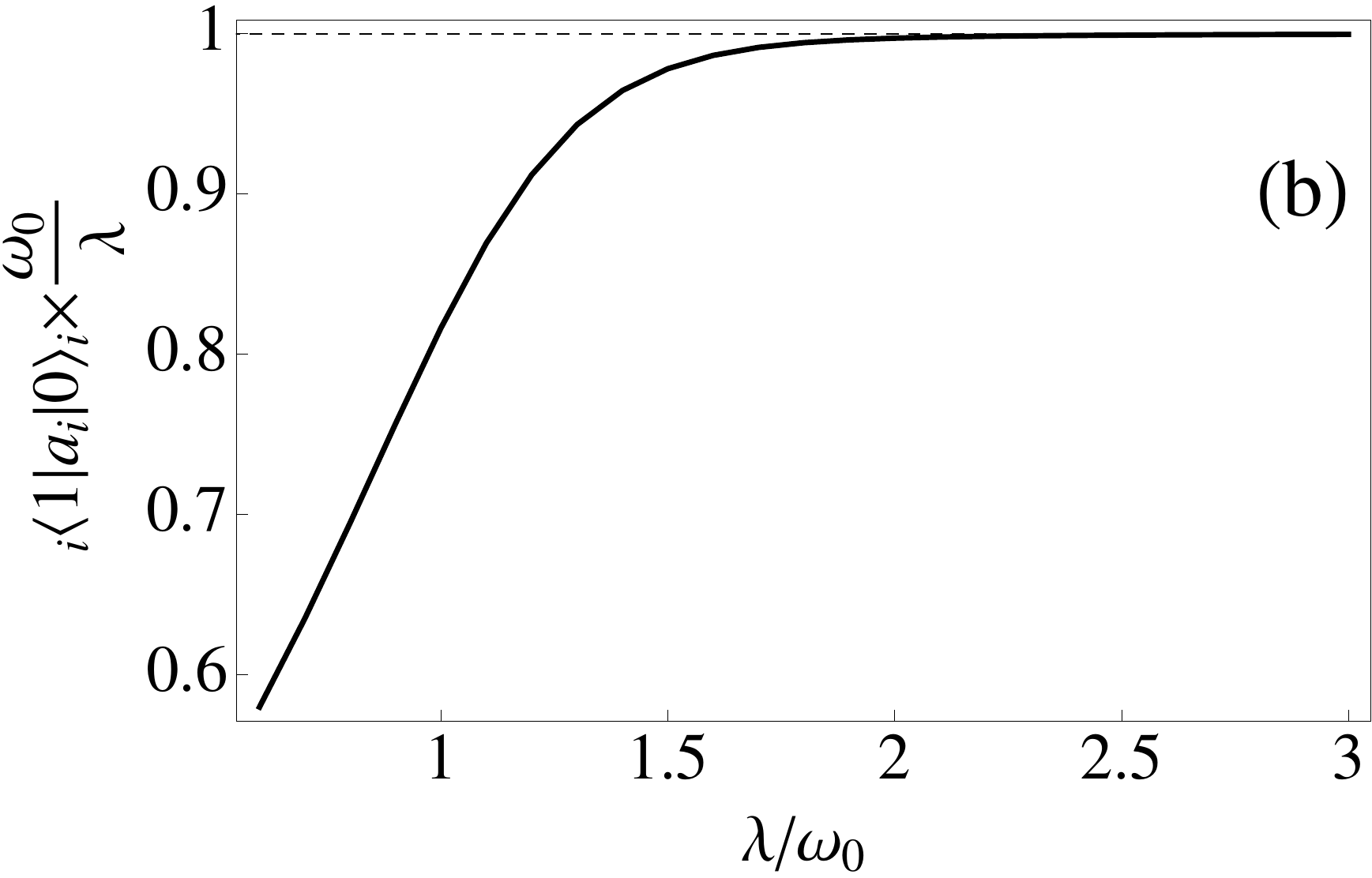}
\caption{(a) Energy diagram for the circuit-QED Hamiltonian
  (\ref{paper::eq:2}) as a function of $\lambda/\omega_0$. (b) Plot of
  $\frac{\omega_0}{\lambda} {}_i{\bra{1}}a_i\ket{0}_i$. We can
  conclude that $\lambda>2\omega_0$ is required for our model to be
  valid because the transverse field $\Delta$
  almost vanishes and the identification of photon annihilation
  operator as a spin flip operator, $\frac{\omega_0}{\lambda}
  a_i=\tau_i^x$, is justified.}
\label{fig:2}
\end{figure}

\emph{Detection and control} --- In this section, we suggest a scheme to
control and measure the non-local spin qubit. It can be also interpreted as
detecting the Majorana bound states. Our proposal consists only of an
additional empty resonator coupled to the resonator at the end of the
circuit-QED chain. Consider a resonator with a frequency, $\omega_d$,
capacitively coupled to $N$th cavity, so that we have
\begin{equation}
\label{paper::eq:11}
H_d=J_d(a_N^\dagger a_d +a_N a_d^\dagger)+\omega_da^\dagger_d a_d
\end{equation}
where $a_N$ represents the field operator of $N$th cavity, and $a_d$ the field
operator of the detection cavity. As shown earlier, the $N$th cavity's
creation and annihilation operators are equivalent to
$\lambda/\omega_0\tau_N^x$ for the $N$th effective spin. Moreover, for the
non-local spin qubits, $\tau_i^x$ is equivalent to
$S^x=\ket{\Psi_0}\bra{\Psi_1}+\ket{\Psi_1}\bra{\Psi_0}$ for any $i$ as
$\tau_i^x\ket{\Psi_s}=\ket{\Psi_{1- s}}$ ($s=0,1$). Therefore, assuming that
$\tilde{J}_d\equiv J_d\lambda/\omega_0\ll J_\mathrm{eff}$, the low-energy
effective Hamiltonian (\ref{paper::eq:7}) combined with the detection
Hamiltonian (\ref{paper::eq:11}) leads again to the Rabi Hamiltonian
\begin{equation}
\label{paper::eq:13}
H_\mathrm{Rabi}
= \frac{\delta}{2}S^z+\tilde J_dS^x(a_d +a_d^\dagger)+\omega_da^\dagger_d a_d
\end{equation}
Here we can make the rotating wave approximation, then the Hamiltonian reduces
to the Jaynes-Cummings Hamiltonian. Therefore, by just adding an empty
resonator at one end of the circuit-QED array, we can realize a circuit-QED
Hamiltonian for the non-local spin qubit. It allows us to tap into the
standard techniques available for the circuit-QED to control and measure the
non-local spin qubit. For example, since the detuning between the detection
cavity frequency and the non-local spin qubit splitting,
$\Delta_d=\omega_d-\delta$ is large compared to $\tilde J_d$, it is in the
dispersive regime where the cavity frequency pulling by the non-local spin
qubit is $ \delta\omega_d={2J_d^2}/{\Delta_d}$~\cite{Blais:2004kn}. This can
be experimentally measured since one can have $\tilde J_d\sim10^{-4}\omega_0$
which achieves the standard strong-coupling regime for the circuit
QED~\cite{Wallraff:2004dy}.

\emph{Experimental feasibility} --- Finally we examine the experimental
feasibility of the ideas explained above, estimating possible values of
physical parameters of the system. Two requirements must be satisfied: First,
the two ground states of each cQED in the system must be nearly degenerate and
well separated from higher excitations. In Fig.~(\ref{fig:2}) (a) are plotted
the energies of individual circuit-QED Hamiltonian (\ref{paper::eq:2}) in the
resonant case ($\omega_0=\Omega$). Figure~\ref{fig:2} (b) plots
$\frac{\omega_0}{\lambda} {}_i{\bra{1}}a_i\ket{0}_i$ to illustrate how good
(its value close to 1) the approximation $a_i=\lambda/\omega_0\tau^x_i$
is. One can see that $\lambda\sim2\omega_0$ suffices for the requirement.
Second, the system should be in the magnetically ordered phase (in terms of
the effective TFIM), $\Delta<J_\mathrm{eff}$ or equivalently
\begin{math}
\Omega\exp\left[-2(\lambda/\omega_0)^2\right] < 4J(\lambda/\omega_0)^2.
\end{math}
This requirement is satisfied provided that $J>10^{-5}\omega_0$. The desired
coupling strength, $\lambda>2\omega_0$, seems achievable for the Fluxonium
coupled inductively to the superconducting
resonator~\cite{Nataf:2011ff}. Moreover, $J>10^{-5}\omega_0$ is also realistic
for the superconducting resonators, with $J$ in the range of a few
MHz.

\emph{Conclusion} -- We have found several intriguing properties of the two
nearly degenerate ground states of a chain of coupled circuit-QED systems in
the ultrastrong coupling regime. The ground states are Schr\"odinger cat
states at a truly large scale, and are mathematically equivalent to Majorana
bound states. With a suitable design of the system, they are protected against
local fluctuations, and may be probed and manipulated coherently by attaching
an extra empty resonator.

Finishing this work, we have noticed a closely related
preprint~\cite{1205.3083S}. While they focus on the phase transition of the
circuit-QED chain, we are mainly concerned about the quantum properties of the
nearly degenerated ground states on one side of the phase transition. In this
respect, both works are complementary to each other.

\bibliographystyle{apsrev4-1}
\bibliography{paper}

\end{document}